\documentclass[%
 reprint,
 superscriptaddress,
 amsmath,amssymb,
 aps,
 pra
]{revtex4-2}

\usepackage{graphicx}
\usepackage{epstopdf}
\usepackage{dcolumn}
\usepackage{bm}
\usepackage{braket}
\usepackage{color}
\usepackage{mathrsfs}
\usepackage{ragged2e}
\usepackage{eqnarray}
\usepackage{subeqnarray}
\usepackage{cases}
\usepackage{amsfonts}
\usepackage{dcolumn}
\usepackage{bm}
\usepackage{tikz}
\usepackage{float}

\usepackage[colorlinks=true,citecolor=blue,urlcolor=blue]{hyperref}
\usepackage{amsmath,amsfonts,amssymb,times,natbib}
\usepackage[standard]{ntheorem}

\begin{document}

\preprint{APS/123-QED}

\title{Long-lived storage of orbital angular momentum quantum states}

\author{Ying-Hao Ye}
\thanks{These authors contributed equally to this work.}
\affiliation{Key Laboratory of Quantum Information, University of Science and Technology of China, Hefei, Anhui 230026, China.
}%
\affiliation{CAS Center for Excellence in Quantum Information and Quantum Physics, University of Science and Technology of China, Hefei, Anhui 230026, China.
}%
\affiliation{Institute for Quantum Control and Quantum Information and School of Physics and Materials Engineering, Hefei Normal University, Hefei, Anhui 230601, China}%

\author{Lei Zeng}
\thanks{These authors contributed equally to this work.}
\affiliation{Key Laboratory of Quantum Information, University of Science and Technology of China, Hefei, Anhui 230026, China.
}%
\affiliation{CAS Center for Excellence in Quantum Information and Quantum Physics, University of Science and Technology of China, Hefei, Anhui 230026, China.
}%

\author{Ming-Xin Dong}
\affiliation{Key Laboratory of Quantum Information, University of Science and Technology of China, Hefei, Anhui 230026, China.
}%
\affiliation{CAS Center for Excellence in Quantum Information and Quantum Physics, University of Science and Technology of China, Hefei, Anhui 230026, China.
}%
\affiliation{Institute for Quantum Control and Quantum Information and School of Physics and Materials Engineering, Hefei Normal University, Hefei, Anhui 230601, China}%

\author{Wei-Hang Zhang}
\author{En-Ze Li}
\affiliation{Key Laboratory of Quantum Information, University of Science and Technology of China, Hefei, Anhui 230026, China.
}%
\affiliation{CAS Center for Excellence in Quantum Information and Quantum Physics, University of Science and Technology of China, Hefei, Anhui 230026, China.
}%

\author{Da-Chuang Li}
\affiliation{Institute for Quantum Control and Quantum Information and School of Physics and Materials Engineering, Hefei Normal University, Hefei, Anhui 230601, China}%

\author{Dong-Sheng Ding}
\email{dds@ustc.edu.cn}
\author{Guang-Can Guo}
\email{gcguo@ustc.edu.cn}
\author{Bao-Sen Shi}
\email{drshi@ustc.edu.cn}
\affiliation{Key Laboratory of Quantum Information, University of Science and Technology of China, Hefei, Anhui 230026, China.
}%
\affiliation{CAS Center for Excellence in Quantum Information and Quantum Physics, University of Science and Technology of China, Hefei, Anhui 230026, China.
}%

\date{\today}


\maketitle

\renewcommand{\thefootnote}{\fnsymbol{footnote}}
\footnotetext[1]{These authors contributed equally to this work.}
\textbf{Quantum memories are indispensible for establishing a long-distance quantum network. High-dimensional quantum memories enable a higher channel capacity compared to a quantum memory working in a two-dimensional space, and have a lower requirement for storage lifetime in the field of quantum coomunication. The photonic transverse spatial modes such as Laguerra-Gaussian modes orbital angular momentum (OAM) are ideal candidates for encoding high-dimensional information, because it can form an infinite-dimensional Hilbert space. Although the faithful storage of an OAM qubit or qutrit has been realized in pioneering works, the longest storage lifetime for the former is only in the order of a few microseconds, and hundreds of nano-seconds for the latter. Here we implement a quantum memory for OAM qubits and qutrits using a cold atomic ensemble, the experimental results clearly show that our memory can still beat the classical limit after a storage time of $400\mu s$,which is two orders of magnitude higher than the previous work. The retrieval efficiency at this time equals to $44\%$ of the value when the storage time is set to be $10\mu s$. Our work is very promising for establishing a high dimensional quantum network.}

Quantum memories that can realize a faithful and reversible transfer of quantum information between photons and matter\cite{opticalmemoryreview,quantummemorylukind} have found important applications in the field of quantum information science such as quantum communication\cite{quantumentanglement,quantumchannelandmemory} and quantum computing. For example, quantum memories are the essential building blocks of quantum repeaters\cite{quantumrepeaterreview}, which is indispensible for a long-distance quantum entanglement distribution. The dimensionality and the storage lifetime are two important performance metrics for a quantum memory, as the storage of higher-dimensional states offers the promise of a significantly improved channel capacity and entanglement generation rate\cite{multimoderepeater}, of easing the requirement on storage time\cite{easestoragelife}, while a longer storage lifetime enables a higher entanglement distribution rate and longer communication distance. Moreover, a quantum key distribution based on higher-dimensional system is more robust to intercept/resend attack\cite{increasederrorrateforeavs} by an eavesdropper.

When it comes to high-dimensional optical memories, a lot of developments have been seen in recent years on the topic of storing information encoded on one or multiple photonic degrees of freedom such as polarization and spatial mode\cite{zhangweinc2016}, path and spatial mode\cite{spatialmultiplex},path and time bin\cite{timebinshanxi,duanluming2020prl,duanluming2017nc}, etc. Among them, the transverse spatial multimode has received widespread attention, because it can be directly stored in an ensemble without the requirement for complex setup. A complete set of solutions of the paraxial Helmholtz equation, such as Laguerre-Gaussian ($\mathrm{LG}$) modes form an infinite-dimensional Hilbert space\cite{LGandHG} that one can span arbitrary transverse mode on, and the capacity of data transmission can therefore be increased to terabit by employing multiplexd $\mathrm{LG}$ modes\cite{terabitOAM}. What's more, a memory for transverse multimode may also find novel applications such as quantum holographic-based teleportation\cite{holographiccommunication} or computing\cite{holographiccomputing}. $\mathrm{LG}$ modes with nonzero azimuthal index ($l \ne 0$) carry a quantized orbital angular momentum $l \hbar$\cite{OAMreview}, thus for the sake of simplicity we refer to a subset of $\mathrm{LG}$ modes with zero radial index ($p=0$) as orbital angular momentum (OAM) modes, and the stored qudits in this article are encoded on these OAM modes.

\begin{figure*}[htb]
	\centering
	\includegraphics[width=500pt]{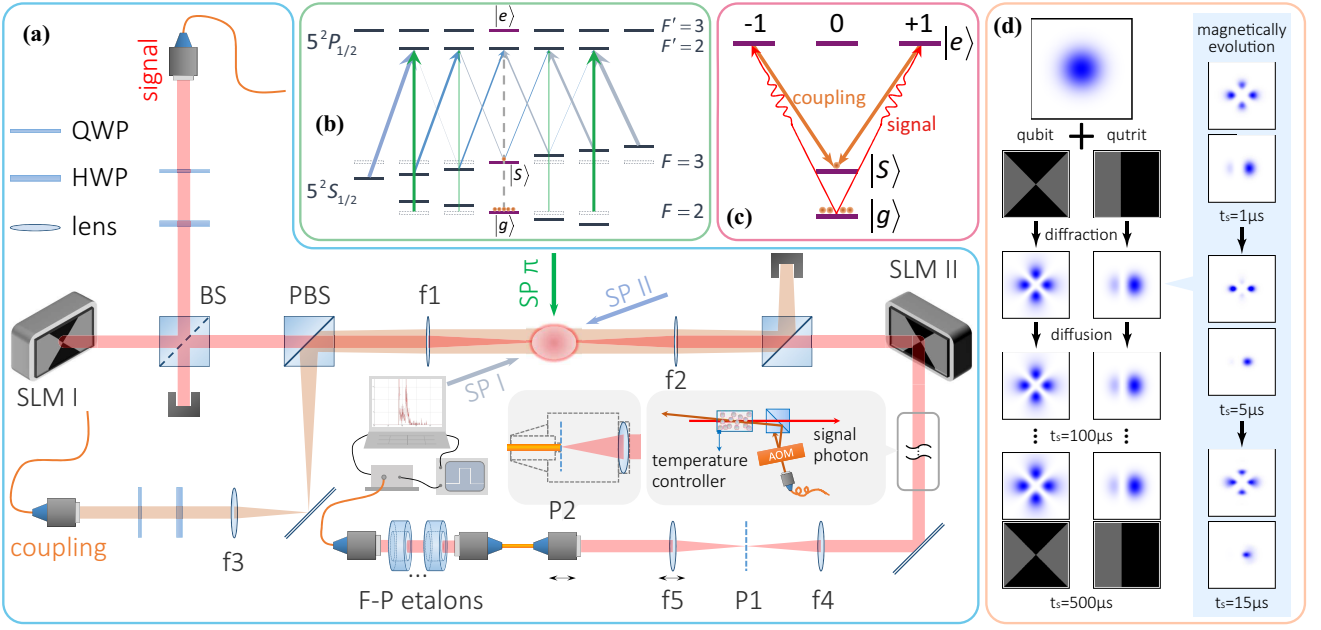}
	\caption{\label{fig:setup} (Color online). The principle of our experiment. (a) Experimental setup. BS: beam splitter; PBS: polarizing beam splitter; SLM: spatial light modulator; QWP: quarter-wave plate; HWP: half-wave plate; AOM: acoustic-optical modulator. (b) Schematic of the clock-state preparation. (c) Associated energy levels for EIT-based optical storage. (d) The illustration of the two main decoherence mechanisom discussed in this article: unsynchronized evolution of stored spin wave caused by the inhomogenous ambient magnetic field and the free expansion of the atomic ensemble. The ambient magnetic field is set to be $5\%$ of the trapping field and the temperature of the ensemble is set to be $100\mu K$.}
\end{figure*}

Since the ground-breaking work of Ref.~\cite{topologicalstability07} has demonstrated that the topological phase singularity of an optical vortex (OAM mode with $\left|l\right|=1$) can resist the strong diffusion of atomic vapor, many works have realized a faithful storage of transverse pattern utilizing artificial\cite{structuredbeams,artificialphase} or intrinsic phase structures\cite{imagestoragevaportheory,imagestoragevaporexp,imagestoragehotvaportheo,multiimagefreconverding}. The storage of OAM qubit carried by a heralded single photon and attenuated single photon have been realized by Bao-Sen Shi's group\cite{ding2013single} and Laurat's group\cite{OAMqubitlaurat} respectively, the former achieved a storage lifetime of about $1\mu s$, the latter achieved a storage time of several microseconds. Subsequently, the dimensionality of quantum memory has been extended to three by Bao-Sen Shi's group\cite{qutritding}, and the storage time for OAM qutrits\cite{qutritding} is on the orders of $10^2ns$. In addition, the storage of high-dimensional OAM states entanglement has also been realized\cite{OAMentanglementstorageding,qutritding2016light}. In this work, we extend the lifetime of stored qudit by not only utilizing the phase structure that is intrinsically robust to the atomic motion, but also using a guiding magnetic field to synchronize the evolution of the atomic coherence and preparing atoms in the magnetically insensitive states. By taking into account these mechanisms, we achieve a storage time exceeding hundreds of microseconds for a qudit carried by pesudo-single photon. The retrieved fidelity of our memory exceeds the best value of a classical memory at least for a storage time ($t_s$) of $400\mu s$. The retrieval efficiency at this time equals to $44\%$ of the retrieval efficiency when $t_s$ is set to be $10\mu s$, which promise the potential applications in quantum region. It is noteworthy to mention that, to the best of our knowledge, the achieved storage lifetime is two orders of magnitude longer than the previous work when storing a qutrit.

\section{\label{sec:level1}Experimental setup and results}
\subsection{Principle of the experiment}
Our memory is based on electro-magnetically induced transparency (EIT) protocol\cite{EITreview} and is implemented in a cold $^{85}\mathrm{Rb}$ ensemble trapped in a 3D magneto-optical trap (MOT). The experimental configuration is shown in Fig.\ref{fig:setup}(a), where the lens $f_1$ and $f_2$ with focal length of $\mathrm{500mm}$ form a $4f$ imaging system that images SLM1 onto SLM2. The signal light from a single mode fiber (SMF) has a $1/e^2$ diameter of $\mathrm{2.1mm}$. The focal plane of the lens $f_1$ coincides with the center of the MOT, therefore the Fraunhofer diffraction of phase pattern imposed on the signal light by the SLM I is stored in the ensemble. To ensure the flatness of the coupling light's transverse distribution in the overlapping area with signal light, a lens $f_3$ with a focal length of $\mathrm{300mm}$, together with the lens $f_1$, form a telescope system that enlarges the waist of coupling beam to $\mathrm{3.2mm}$, therefore we can treat the coupling light as plane wave with unity transversal intensity in the MOT. When considering the projection measurement, the lens $f_4$ transforms the transverse pattern of the signal light right after the projector SLM II into its Fourier spectrum (in the plane $P1$). The lens $f_5$ and the collimation package (F220FC-780 from Thorlabs, with a focal length of $\mathrm{11.07mm}$) form a $4f$ imaging system that images the phase pattern on the $P1$ on to the fiber tip $P2$ to achieve a higher interference visibility. The angle between incident light and reflected light on the SLM II (5 degrees in practical) has been exaggerated in Fig.~\ref{fig:setup} (a) for clarity.

A filtering system consisting of a Glan-Taylor prism, a rubidium absorption cell and three Fabry–Pérot etalons is utilized to filter out the strong co-propagating coupling light from the weak signal light. The absorption cell is heated to $65^{\circ}\mathrm{C}$, a strong ($\mathrm{20mW}$) and expanded pump beam that is on resonance with the signal light is emitted into the cell and counter-propagates with the signal light (see the inset of Fig.\ref{fig:setup}(a)) to pump the atoms into $\left|5S_{1/2}, F=3\right\rangle$. This pump light is switch off by an acousto-optic modulator at the retrieval window to diminish the system noise.

An effective method to eliminate the detrimental effect of the ambient magnetic field on the storage lifetime is mapping photonic information onto the coherence of magnetically insensitive states, because the latter are eigenstates of interaction Hamiltonian with eigenvalue zero. The magnetically insensitive states we employed are the clock states of $^{85}\mathrm{Rb}$, namely, $\left|5S_{1/2}, F=2, m_{F}=0\right\rangle$ (denoted as $\left|g\right\rangle$) and $\left|5S_{1/2}, F=3, m_F=0\right\rangle$ (denoted as $\left|s\right\rangle$). The signal light addresses the transition between $\left|g\right\rangle$ and the excited state $\left|5P_{1/2},F'=3,m_{F'}=0\right\rangle$ (denoted as $\left|e\right\rangle$) while the the coupling light of Rabi frequency $\Omega_c$ resonantly drive the $\left|s\right\rangle\leftrightarrow\left|e\right\rangle$ transition. During each experimental cycle, the atoms are prepared into $\left|g\right\rangle$ initially, this goal can be achieved by utilizing the dipole-forbidden transition\cite{rubidium85data} between $\left|g\right\rangle$ and $\left|e\right\rangle$ as illustrated in Fig.\ref{fig:setup}(b). At first, a strong and uniform polarization magnetic field ($\mathrm{0.97G}$) that is in parallel with the signal light is established in the vicinity of the ensemble by a pair of Helmholtz coils and remains on during the experimental window. Then a beam of high-frequency $\pi$-polarized state preparation light (labelled as SP $\pi$) that is on resonance with the $\left|5S_{1/2}, F=2\right\rangle\leftrightarrow\left|5P_{1/2}, F'=2\right\rangle$ transition and two beams of mutually orthogonal circularly polarized low-frequency state preparation light (labelled as SP I and SP II) are incident on the ensemble simultaneously for $\mathrm{30}\mu\mathrm{s}$. The SP I and SP II light addressing the $\left|5S_{1/2}, F=3\right\rangle\leftrightarrow\left|5P_{1/2}, F'=2\right\rangle$ transition are blue-detuned and red-detuned from the $\left|m_{F}=0\right\rangle\leftrightarrow\left|m_{F'}=0\right\rangle$ by $\mathrm{1.8MHz}$ respectively. A higher atomic population in the desired state can be achieved by using two beams of detuned low-frequency state preparation light instead of one resonant light\cite{statepreparation}. The preparation efficiency of our experiment is measured to be $\mathrm{71}\%$. The associated energy levels for an optical memory based on EIT protocol are shown in Fig.\ref{fig:setup}(c), the weak signal pulse and strong coupling pulse are orthogonally linearly polarized.

\begin{figure*}[htb]
	\centering
	\includegraphics[width=400pt]{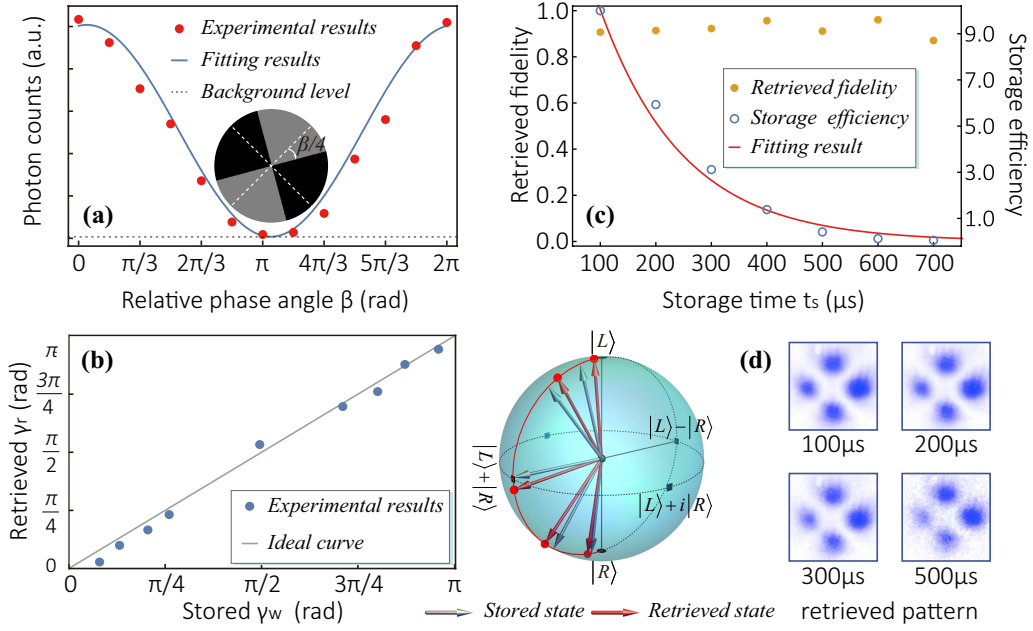}\caption{\label{fig:qubitresult} (Color online) Experimental data for storing OAM qubits. (a) The interference curve achieved by measuring a fixed qubit with basis rotating along the equator of the Bloch sphere; (b) The weight of different OAM components in retrieved qubit $\gamma_r$ as a function of stored $\gamma_w$. (c) The decay of retrieved fidelity and storage efficiency with storage time. (d) The integrated intensity of retrieved pattern recorded by an ICCD, the first row shows the results of different storage time and the second row shows the results of different $l$ at a fixed $t_s=300\mu s$.}
\end{figure*}

To account for the inhomogeneity present in the transverse cross section of the signal light, here we introduce the slowly-varying spatially-dependent signal field annihilation operator $\tilde{\mathcal{E}}\left(z,\vec \rho,t\right)=\hat{\mathcal{E}}\left(z,\vec{\rho},t\right)\mathrm{exp}\left(-ik_{s}z+i\omega_s t\right)$ and atomic operator $\tilde{\sigma}_{mn}\left(z,\vec{\rho},t\right)=\hat{\sigma}_{mn}\left(z,\vec{\rho},t\right)\mathrm{exp}\left[i\left(\omega_{mn}/c\right)t-i\omega_{mn}t\right]$, where $\vec{\rho}=\vec{\rho}\left(x,y\right)$ denotes the transverse coordinate, $k_s$ and $\omega_s$ are the average wave vector and center frequency of the quasi-monochromatic signal light, $\omega_{mn}$is the atomic resonance frequency and $\tilde{\sigma}_{mn}\left(z,\vec{\rho},t\right)=\left(1/N_z\right)\sum\nolimits_{j=1}^{N_z}{\left|m_j\right\rangle\left\langle n_j\right|}$ is the average of flip operator for the $N_z$ atoms in a small but macroscopic volumes at $\left(z,\vec \rho\right)$. The quantization axis is chosen to be in parallel with the signal light ($z$ axial) and $\Omega_c$ is assumed to be real and spatially homogeneous. Because the Rabi frequency of signal light is much smaller than $\Omega_c$ and the number density of photons of input signal pulse is much smaller than the atomic number density, we can follow the method proposed in Ref.~\cite{DSPlukind} and derive the following closed evolution equations for $\tilde{\mathcal{E}}$ and $\tilde{\sigma}_{mn}$ under paraxial and adiabatic approximation:

\begin{subequations}
\label{eqn:1}
\begin{eqnarray}
&&\left(\frac{\partial}{\partial t}+c\frac{\partial}{\partial z}-\frac{ic}{2k_s}\Delta_{\bot}\right)\tilde{\mathcal{E}}\left(z,\vec{\rho},t\right)\nonumber\\
&&=\frac{gN}{\Omega_c\left(t\right)} e^{i\Delta kz}\frac{\partial}{\partial t}\tilde{\sigma}_{gs}\left(z,\vec{\rho},t\right);
\label{eqn:1a}
\end{eqnarray}
\begin{equation}
\tilde{\sigma}_{gs}\left(z,\vec{\rho},t\right)=-g\frac{\tilde{\mathcal{E}}\left(z,\vec{\rho},t\right)}{\Omega_c\left(t\right)}e^{-i\Delta kz},
\label{eqn:1b}
\end{equation}
\end{subequations}
Where $\Delta_\bot=\partial_{\vec{\rho}}^2=\partial_x^2+\partial_y^2$ represents the transverse Laplace operator. $\Delta k=k_c\left(\rm{cos}\alpha-1\right)$ represents the photon momentum that is transferred to the atomic ensemble, where $\alpha$ is the angle between coupling light and signal light, therefore we refer to $\Delta k$ as the wave vector of  the spin coherence in the following. To solve the equation analytically we perform two-dimensional Fourier transform as with the Ref.~\cite{multimoderamantheory}:
\begin{equation}
\tilde{\mathbb{E}}\left(z,t;\vec{q}\right)=e^{-iq^2z/\left(2k_s\right)}\frac{1}{2 \pi}\iint{\tilde{\mathcal{E}}\left(z,\vec{\rho},t\right)e^{-i\vec{q}\cdot\vec{\rho}}}\rm{d}^2\rho,
\label{eqn:2}
\end{equation}
\begin{equation}
\tilde{\mathbb{\sigma}}_{mn}\left(z,t;\vec{q}\right)=e^{-iq^2z/\left(2k_s\right)}\frac{1}{2 \pi}\iint{\tilde{\sigma}_{mn}\left(z,\vec{\rho},t\right)e^{-i\vec{q}\cdot\vec{\rho}}}\rm{d}^2\rho.
\label{eqn:3}
\end{equation}
By substituting Eq.~(\ref{eqn:2}) and Eq.~(\ref{eqn:3}) into Eqs.~(\ref{eqn:1}) we obtain the propagation of the signal field:
\begin{equation}
\left(\frac{\partial}{\partial t}+c\frac{\partial}{\partial z}\right)\tilde{\mathbb{E}}\left(z,t;\vec{q}\right)=-\frac{g^2N}{\Omega_c\left(t\right)}\frac{\partial}{\partial t}\frac{\tilde{\mathbb{E}}\left(z,t;\vec q\right)}{\Omega_c\left(t\right)}.
\label{eqn:4}
\end{equation}
The above equation is identical to the equation in Ref.~\cite{DSPlukind} except that it is in the Fourier domain. Introducing the following quantum states:
\begin{eqnarray}
\hat{\Psi}\left(z,t;\vec{q}\right)=&&\mathrm{cos}\theta\left(t\right)\tilde{\mathbb{E}}\left(z,t;\vec{q}\right)\nonumber\\
&&-\mathrm{sin}\theta\left(t\right)\sqrt{N}\tilde{\sigma}_{gs}\left(z,t;\vec{q}\right)e^{i\Delta kz};
\label{eqn:5}
\end{eqnarray}
\begin{eqnarray}
\hat{\Phi}\left(z,t;\vec{q}\right)=&&\mathrm{sin}\theta\left(t\right)\tilde{\mathbb{E}}\left(z,t;\vec{q}\right)\nonumber\\
&&+\mathrm{cos}\theta\left(t\right)\sqrt{N}\tilde{\sigma}_{gs}\left(z,t;\vec{q}\right)e^{i\Delta kz}.
\label{eqn:6}
\end{eqnarray}
Clearly $\hat{\Psi}$ has no contribution from $\left|e\right\rangle$ and is the eigenstate of interaction Hamiltonian between light and atoms with eigenvalue zero, therefore it is also called dark state. What’s more, this dark state for a single plane-wave components obeys bosonic commutation relations $\left[\hat{\Psi}_k\left(z,t;\vec{q}\right),\hat{\Psi}_{k'}^\dag\left(z,t;\vec{q}\right)\right]\approx\delta_{k,k'}$, so that we can associate it with a bosonic quasi-particle that called dark-state polariton (DSP). By substituting Eq.~(\ref{eqn:5}) and Eq.~(\ref{eqn:6}) into Eq.~(\ref{eqn:4}) and taking into account the adiabatic approximation ($\hat{\Phi}=0$), one can achieve a rather simple propagation equation for $\hat\Psi$:
\begin{equation}
\left[\frac{\partial}{\partial t}+c\mathrm{cos}^2\theta\left(t\right)\frac{\partial}{\partial z}\right]\hat\Psi\left(z,t;\vec{q}\right)=0.
\label{eqn:7}
\end{equation}
Where $\mathrm{tan}^2\theta\left(t\right)=g^2N/\Omega_c^2\left(t\right)$. The above equation describes a transverse distribution and quantum-state preserved propagation of DSPs with an instantaneous group velocity $v_g=c\mathrm{cos}^2\theta\left(t\right)=c/\left(1+g^2N/\Omega _c^2\right)$. At the beginning of a storage process, the strong coupling light satisfies $\Omega_c^2\gg g^2N$ and thus $\theta\rightarrow 0$, the signal pulse enters the front face of atomic ensemble and propagates in the medium with a group velocity $v_g$ approaches $c$. By switching off the coupling light adiabatically, the mixing angle $\theta$ is changed to $\pi/2$ and DSPs are decelerated to $v_g=0$, during this process, the electromagnetic part of the DSP is transferred into the long-lived ground states coherence of rubidium atoms with its transverse intensity and phase distribution preserved:
\begin{equation}
\tilde{\mathcal{E}}\Leftrightarrow-\sqrt{N}\mathcal{F}^{-1}\left\{\mathrm{exp}\left(-i\frac{q^2z'}{2k_s}\right)\mathcal{F}\left[\tilde\sigma_{gs}\left(z,\vec \rho,t\right)\right]\right\}e^{i\Delta kz'},
\label{eqn:8}
\end{equation}
where $\mathcal{F}\left[\cdots\right]$ denotes Fourier transform. The experimental parameters satisfy that $q^2z'/2k_s\le q^2D/k_s\ll 1$, where $D$ is the diameter of the atomic ensemble and is measured to be 2mm, therefore we omit the diffraction factor $\mathrm{exp}\left(-iq^2z'/2k_s\right)$ in the following analysis and we have approximately:
\begin{equation}
\tilde{\mathcal{E}}\left(0,\vec{\rho},t\right)\Rightarrow -\sqrt{N}\tilde{\sigma}_{gs}\left(z',\vec{\rho}\right)e^{i\Delta kz'}.
\label{eqn:9}
\end{equation}
This equation is identical to the case of storing fundamental Gaussian modes. In fact, it has been proved theoretically that the diffraction effect in the writing process can be compensated in the reading process by using a forward readout setup\cite{multimoderamantheory}, therefore the above approximation can be concerned as justified. From Eq.~(\ref{eqn:9}) it is obvious that the spin coherence will be quite vulnerable to the drifting of atoms in the longitude direction when the spin wave vector is large, thus a small $\Delta k$ is desirable to diminish the detrimental effect of atomic motion. In this experiment we let the signal light and coupling light to be collinear to reduce the wave vector of spin wave\cite{colinearmillisecond}.
\begin{figure*}[htb]
	\centering
	\includegraphics[width=450pt]{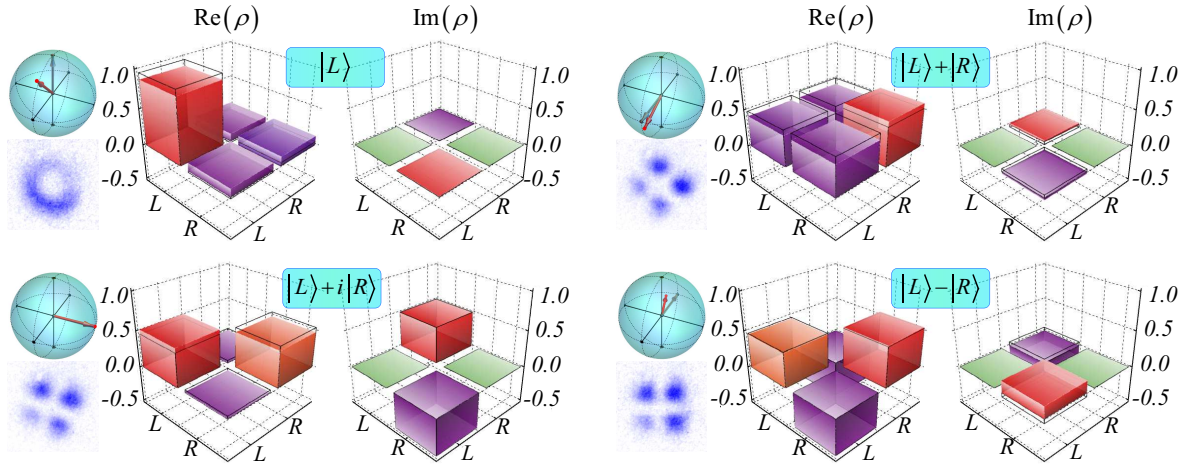}\caption{\label{fig:qubittomo} The reconstructed density operators of different OAM qubits after $500\mu s$ of storage, mean number of photons per pulse here is 50 and the result here has subtracted background noise. The black border represents the input value, the retrieved pattern, the locations of ideal and retrieved state in the Bloch sphere are depicted on the left hand side of each tomography.}
\end{figure*}

There are two important mechanisms contribute to the decoherence of transverse multimodes, the atomic diffusion in the transverse direction and the dephase effect caused by ambient residual magnetic field. When it comes to the diffusion, it is worth mentioning that the superposition states of Laguerre-Gaussian modes stored in the ensemble are intrinsically robust to the transverse atomic motion due to the destructive interference in the dark region between two adjacent bright area. This effect can be illustrated by the example in Fig.\ref{fig:setup} (d), the stored qubit $\left|L\right\rangle+\left|R\right\rangle$ and qutrit $\left|L\right\rangle+\left|G\right\rangle+\left|R\right\rangle$ (where $\left|L\right\rangle$, $\left|R\right\rangle$ and $\left|G\right\rangle$ stand for LG states with OAM $+l$, $-l$ and $0$ respectively) are generated by Fraunhofer diffracting the phase structure $\mathrm{Arg}\left[\mathrm{cos} 2 \phi\right]$ and $\mathrm{Arg}\left[1+2\sqrt{2}r\mathrm{cos}\phi/\omega_0\right]$ imprinted on a fundamental Gaussian mode through the SLM I respectively\cite{diffractiontheory}, where $r$ and $\phi$ are the radial coordinate and azimuthal angle in the transverse plane, $\mathrm{Arg}\left[...\right]$ indicates taking the argument. It's obvious that the phase of hologram on the SLM I can only take values $0$ or $\pi$, thus there will always exist a flipped phase between two bright area after the diffraction, as can be seen in Fig.\ref{fig:setup} (d). The initial collective coherence distribution is proportion to the input signal field as Eq.~\ref{eqn:9}, when considering the free expansion of the ensemble, the diffused distribution can be calculated by performing the convolution\cite{imagestoragevaportheory}:
\begin{eqnarray}
\tilde{\sigma}_{gs}\left(z',\vec{\rho},t_s\right)=&&\frac{m}{2\pi k_B T t_s^2}\iint{\tilde{\sigma}_{gs}\left(z',\vec{\rho'},t_s=0\right)}\nonumber\\
&&\times
\mathrm{exp}\left(-m\frac{\left|\vec{\rho}-\vec{\rho'}\right|^2}{2k_B T t_s^2}\right)\mathrm{d}^2\vec{\rho'}
\label{eqn:10}
\end{eqnarray}
where $k_B$ is the Boltzmann constant and $T$ is the absolute temperature of the ensemble, the latter is estimated to be $100\mu K$ in a independent time-of-flight experiment. Eq.~\ref{eqn:10} assumes a Gaussian distribution of number density of atoms in the ensemble. The simulated results for diffused intensity and phase pattern are shown in Fig.\ref{fig:setup}(d). It is clear that the intensity and phase structure are well preserved except for an overall enlargement. Because of the difference in symmetry, the position of dark lines of the qubit remains unchanged while the dark line of the qutrit move to the left after diffusion, therefore we expect a slightly more rapid decreasing trend for retrieved fidelity of a stored qutrit compared to that of a stored qubit.

A much more important mechanism that leads to the decoherence is the inhomogeneous evolution of the spin wave in the transverse plane caused by the ambient magnetic field\cite{synchronizedye}, this is because typically the Larmor period is much shorter than the time scale of other decoherence mechanisms and the unsynchronized precession of atomic magnetic moments in the transverse plane will cause distortion in the retrieved pattern. This effect is illustrated by the simulations presented in Fig.~\ref{fig:setup} (d), the ambient magnetic field is set to be $5\%$ of the trapping field there and we use the method proposed in Ref.~\cite{synchronizedye}. Clearly this unsynchronized evolution dominate the decoherence of stored spin wave at short storage time. To overcome this effect, we use the strong quantization magnetic field for state preparation as a guiding field to synchronize the precession of atomic magnetic moments. What's more, by mapping the photonic information to the coherence of a pair of magnetic insensitive states, the destructive interference between different Zeeman sub-levels is also eliminated\cite{DSPrevival,DSPrevtheory}, therefore we can extend the storage lifetime to the orders of $10^2\mu s$.
\subsection{Experimental results of storing qubits}

To characterize the performance of our optical memory, we store a set of input qubits distributed over the Bloch sphere that carried by a weak coherent signal pulse (the average photon number per pulse $\bar{n}\approx 50$). The OAM qubit to be stored can be generally formulated as $\mathrm{cos}\left(\gamma/2\right)\left|L\right\rangle+\mathrm{sin}\left(\gamma/2\right)e^{i\beta}\left|R\right\rangle$, where the azimuthal angle $\beta$ and the polar angle $\gamma$ of the Bloch sphere determine the relative phase and weight between these two OAM components respectively. First, we store a fixed OAM qubit $\left|L\right\rangle+\left|R\right\rangle$ with $l=2$ for $500\mu s$. After retrieval, we measure the retrieved photons by scanning the measurement base along the equator of the Bloch sphere as $\left|L\right\rangle+e^{i\beta}\left|R\right\rangle$, i.e., by displaying $\left|R\right\rangle+e^{i\beta}\left|L\right\rangle$ on the SLM II, see the inset of Fig.\ref{fig:qubitresult} (a). A gated single photon detector (SPD) registers the photons coupled by the SMF in the retrieval window and the ideal curve of integrated SPD counts versus $\beta$ is proportional to $1+\mathrm{cos}\beta$. We collect a data point for every $\pi/6$ change in $\beta$ and the results are shown as red dots in Fig.\ref{fig:qubitresult} (a), those points are fitted with $N\left(\beta\right)=N_0\left(1+\delta+\mathrm{cos}\beta\right)$, where $\delta$ is a fitting parameter, the result is shown in a blue sold line. A criterion for the deviation of measured curve from the ideal curve is the interference visibility, The achieved visibility is 0.954 and 0.981 without and with background noise subtraction respectively, the latter is close to the value (0.998) of the input signal. We have also measured the visibility of the retrieved OAM qubit with different $l$ and the results are 0.987 and 0.992 for $l=2$ and $l=3$ respectively with background noise correction.

Next, we test the validity of storing and retrieving OAM qubits along a meridian on the Bloch sphere by keeping $\beta$ at zero and scan the value of $\gamma$ from $0$ to $\pi$. Each qubit is stored for $500\mu s$ and the retrieved signal is projected to basis $\left|L\right\rangle$ and $\left|R\right\rangle$. The retrieved polar angle $\gamma_{r}$ can be calculated as $\gamma_r=2\mathrm{arctan}\sqrt{N_R/N_L}$, where $N_R$ and $N_L$ is the integrated counts of the SPD on the basis $\left|L\right\rangle$ and $\left|R\right\rangle$ respectively. The retrieved $\gamma_r$ is plotted as a function of original polar angle $\gamma_w$ in Fig.\ref{fig:qubitresult} (b), $\gamma_r=\gamma_w$ indicates a faithful storage. The experimental data are distributed in the vicinity of the ideal curve (gray solid line) and the deviation variance between measured value and ideal values is calculated to be $0.10$. The corresponding points are also shown on the Bloch sphere in Fig.\ref{fig:qubitresult} (b), where the gray arrows point to the original points and the red arrows point to the measured points.

We have demonstrated a faithful storage of OAM qubits distributed along the longitude and latitude of the Bloch sphere, it’s time for us to reconstruct the density operator of retrieved OAM qubits by performing quantum state tomography (QST), the chosen basis for the projection measurement are $\left|L\right\rangle$, $\left|R\right\rangle$, $\left|L\right\rangle+\left|R\right\rangle$, $\left|L\right\rangle+i\left|R\right\rangle $ and $\left|L\right\rangle-\left|R\right\rangle$. The similarity between the retrieved signal and original signal is quantified by fidelity defined as $F\left(\rho,\rho_0\right)=\mathrm{tr}\left[\rho^{1/2}\rho_0\rho^{1/2}\right]^{1/2}$, where $\rho$ is the density operator of the retrieved state and $\rho_0$ is the density operator of the input state or ideal state. The decoherence of transverse information is indicated by the decay of fidelity between retrieved signal and input signal over storage time $t_s$, see the orange circles in Fig.\ref{fig:qubitresult} (c), the retrieval efficiency as a function of $t_s$ is plotted by empty blue circles and the solid red line is the corresponding fitting result with a exponential function. The fidelity barely decays while the retrieval efficiency exhibits a clear trend of decreasing. To visualize the read-out signal more vividly, we also store a strong coherence signal pulse and capture its retrieved pattern with an ICCD (iStar series from the Andor, all patterns are integrated intensity of 1000 times of captures) , the results at different $t_s$ are listed in Fig.\ref{fig:qubitresult} (d). We then keep $t_s$ at $500\mu s$ and perform QSTs for different OAM qubit states, the results are shown in Fig.\ref{fig:qubittomo} and associated fidelities are tabulated in Table.~\ref{tab:table1}. The black borders represent the input value. The retrieved pattern at $t_s=500\mu s$, the locations of ideal and retrieved state in the Bloch sphere of individual qubit are depicted on the left hand side of each tomography. The second column of Table.~\ref{tab:table1} summarizes the absolute fidelity between readout states and ideal states while the third column summarizes the relative fidelity between readout states and input states.

\begin{table}[h]
\caption{\label{tab:table1}
Fidelities of the retrieved states for different input qubits with $l=2$ at $t_s=500\mu s$, all the results are with background subtraction.
}
\begin{ruledtabular}
\begin{tabular}{lll}
\textrm{Stored qubit}&
\textrm{Absolute fidelity}&
\textrm{Relative fidelity}\\
\colrule
$\left|L\right\rangle$ & 87.22\%\footnote{This relatively low absolute fidelity may caused by the imperfect generation of the state since the relative fidelity of it is not significantly lower than other qubit states'.} & 96.57\%\\
$\left|R\right\rangle$ & 90.21\% & 93.62\%\\
$\left|L\right\rangle+\left|R\right\rangle$ & 91.09\% & 91.16\%\\
$\left|L\right\rangle+i\left|R\right\rangle$ & 99.66\% & 99.48\%\\
$\left|L\right\rangle-\left|R\right\rangle$ & 99.91\% & 97.55\%\\
\end{tabular}
\end{ruledtabular}
\end{table}
\subsection{Experimental results of storing qutrits}
A crucial goal of this article is to extend the number of dimensions that our optical memory are capable of storing to three, i.e., OAM qutrit. In order to get the suitable hologram for generating the desired qutrits, we have to take the device parameters of SLM I (Pluto series from HoloEye, with a resolution of $\mathrm{1920}\times\mathrm{1080}$ and a pixel pitch of $16.0\mu m$) and the diameter of input signal beam into account. The calculated hologram is then fine-adjusted by performing projection measurement with a strong calibration light on the nine basis for tomography: $\left|L\right\rangle$, $\left|G\right\rangle$, $\left|R\right\rangle$, $\left|G\right\rangle-\left|L\right\rangle$, $\left|G\right\rangle+\left|R\right\rangle$, $\left|G\right\rangle+i\left|L\right\rangle$, $\left|G\right\rangle-i\left|R\right\rangle$, $\left|L\right\rangle+i\left|R\right\rangle$ and $\left|L\right\rangle-\left|R\right\rangle$. The atomic MOT and the absorption cell is turned off during the adjustment. In the data processing stage, we shall take into consideration the influence of the non-homogenous absorption of the warm Rb cell on a high-dimensional transverse multimode due to the non-uniform intensity distribution of the pump light even though it has been expanded. This is accomplished by measuring different transmittances of the calibration light in a particular OAM qutrit state after it passing through the absorption cell and projection measurement. The data under $i\mathrm{th}$ projection base is corrected through dividing the raw photon counts by an associated relative transmittance: $T_i^r=T_i/\mathrm{max}\left(T_j\right)$, where $T_{i}$ is the measured transmittance and $\mathrm{max}\left(T_j\right)$ is the maximum transmittance among nine basis.
\begin{figure}[t]
	\centering
	\includegraphics[width=\linewidth]{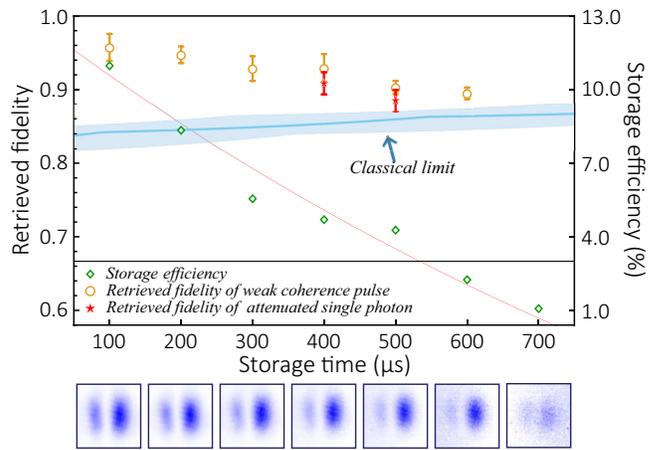}\caption{\label{fig:qutritresult} Time evolution of the stored qutrits. All the data points shown here are the average results of qutrit $\left|L\right\rangle+\left|G\right\rangle+\left|R\right\rangle$ and $\left|L\right\rangle-\left|G\right\rangle+\left|R\right\rangle$, the retrieved fidelity beats the classical limit at $t_s=400\mu s$.}
\end{figure}

At first we store a weak coherence signal pulse with $\overline{n}\approx 50$ carries an OAM qutrit for different $t_s$. The retrieved fidelity as a function of storage time is plotted in Fig.~\ref{fig:qutritresult}, the yellow empty circles correspond to the relative fidelities between read-out states and the input states, all the data have subtracted background noise, and are the average of the results of two qutrit with $l=1$: $\left|\Psi_A\right\rangle=\left|L\right\rangle-\left|G\right\rangle+\left|R\right\rangle$ and $\left|\Psi_B\right\rangle=\left|L\right\rangle+\left|G\right\rangle+\left|R\right\rangle$. The error bars represent the standard deviations of fidelities. As with the case of qubit, the decay of fidelities are much slower than the decay of the retrieval efficiency. The retrieved patterns of $\left|\Psi_B\right\rangle$ are shown at the bottom of Fig.~\ref{fig:qutritresult}, the transverse spatial structure of the input signal is preserved at least for hundreds of milliseconds.

\begin{figure}[htb]
	\centering
	\includegraphics[width=\linewidth]{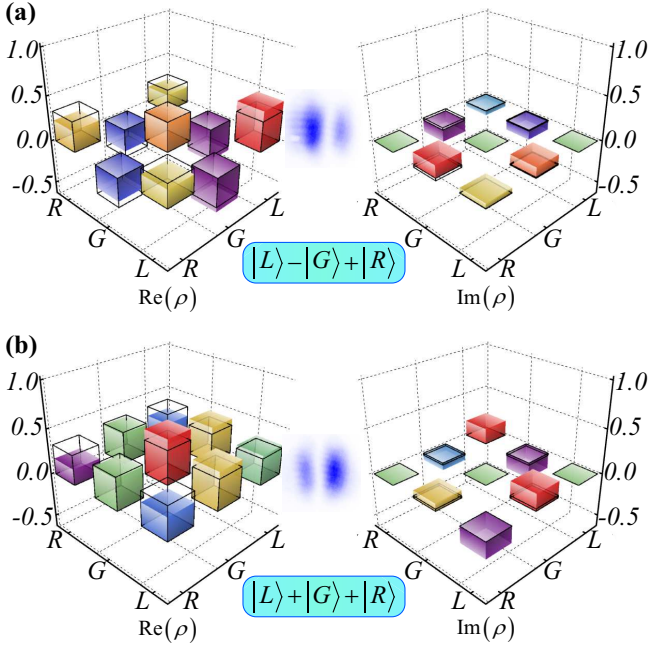}\caption{\label{fig:qutrittomo} The reconstructed density operators of different OAM qutrits after $500\mu s$ of storage. Mean photon number per pulse is $n=1.6\pm 0.4$ (with background noise correction), the black border is the input states, and the retrieved patterns of a strong signal pulse at $t_s=500\mu s$ are also shown here.}
\end{figure}

Next, we further attenuate the signal pulse to single-photon level with $n=1.6\pm 0.4$ to prove the quantum character of our memory, the uncertainty of the mean photon number is estimated by the $3\sigma$ rule. We measured the fidelities of the retrieved signal at $t_s=400\mu s$ and $t_s=500\mu s$, the results are plotted by pentagrams in Fig.~\ref{fig:qutritresult} and tabulated in Table.~\ref{tab:table1}. The associated QSTs of retrieved states at $t_s=500\mu_s$ are depicted in Fig.~\ref{fig:qutrittomo}, where the black boarders indicate the input states. We find that the achieved fidelities are slightly lower than that of the weak coherent signal light. This difference may be due to the longer acquisition time we adopt in the case of attenuated single photon signal pulse, i.e., $\mathrm{1200}s$ for each base, compared to the case of weak coherent signal pulse, where the acquisition time is only $\mathrm{300}s$ for each base. During this long experimental time, the long drift of the system has an increasing impact on the achieved fidelities.

The fidelities of a quantum memory must exceeds the highest value that a classical memory can achieve to guarantee its quantum character. An eavesdropper can utilize a technique called intercept/resend attack to achieve a maximum fidelity of $(N+1)/(N+2)$ for a signal pulse containing $N$ photons\cite{classicallimitearly}. For an ideal single photon, this limits is $2/3$, as shown by the solid black line in Fig.~\ref{fig:qutritresult}. However, an attenuated coherent pulse exhibits the Poissonian statistics $P_{\bar{n}}\left(N\right)=e^{-\bar{n}}\bar{n}^N/N!$, therefore the corrected classical limits should be the weighted average of the maximum achievable fidelity of signal pulses containing different photon number $N$\cite{classicallimitpoisso}:

\begin{equation}
F_{co}\left(\bar{n}\right)=\sum\limits_{N=1}^\infty{\frac{N+1}{N+2}\frac{P_{\bar{n}}\left(N\right)}{1-P_{\bar{n}}\left(0\right)}}.
\label{eqn:11}
\end{equation}
What’s more, the eavesdropper can take advantage of the non-unit retrieval efficiency to increase the fidelity by resending a read-out photon only if the number of photons per pulse is high enough. Therefore the classical limit then reads\cite{classicallimitpoisso,polarizationqubitsolid2}:

\begin{eqnarray}
\begin{aligned}
&F_{classical}\left(\bar n,\eta_m\right)\\
&=\frac{\left(\frac{N_{min}+1}{N_{min}+2}\right)\gamma+\sum\nolimits_{N\ge N_{min}+1}^\infty{\frac{N+1}{N+2}P_{\bar{n}}\left(N\right)}}{\gamma+\sum\nolimits_{N\ge N_{min}+1}{P_{\bar{n}}\left(N\right)}}.
\end{aligned}
\label{eqn:12}
\end{eqnarray}
Where $\gamma=\eta_m\left[1-P_{\bar{n}}\left(0\right)\right]-\sum\nolimits_{N\ge N_{min}+1}{P_{\bar{n}}\left(N\right)}$, $\eta_m$ is the retrieval efficiency and $N_{min}$ is the minimum value that satisfies $\sum\nolimits_{N\ge N_{min}+1}P_{\bar{n}}\left(N\right)\le \left[1-P_{\bar{n}}\left(0\right)\right]\eta_m$. The aforementioned classical limit depends on the memory efficiency, so that we record the retrieval efficiency as a function of storage time and plot it by red empty diamonds in Fig.~\ref{fig:qutritresult}. The fitting results of efficiency is also shown in the picture, see the solid red line. The classical limit is then calculated by substituting the time-dependent retrieval efficiency $\eta_m\left(t_s\right)$ and $\bar{n}=1.6\pm 0.4$ into Eq.~\ref{eqn:12}, see the solid blue line in Fig.~\ref{fig:qutritresult}. The light-blue-shaded region is the threshold band after taking into account the uncertainty of the mean photon per pulse. As can be seem from the figure, the achieved fidelity is well above the upper boundary of the threshold band at $t_{s}=400\mu s$, the retrieval efficiency is $4.73\%$ at this time, which is $44\%$ of that at $t_s=10\mu s$ ($10.74\%$). Therefore our memory can still work in quantum realm at least for a storage time of $400\mu s$.

\begin{table}[htb]
	\caption{\label{tab:table2}
		The retrieved fidelities for different input qutrits with $l=1$ at $t_s=500\mu s$, all the results are with background subtraction. The fourth row is the upper bound of the quantum-classical threshold when considering the uncertainty of mean photon number, the Poissonian nature of the attenuated single photon and non-unit retrieval efficiency.
	}
	\begin{ruledtabular}
		\begin{tabular}{llll}
			\textrm{Stored qutrit}&
			\textrm{Fidelity A\footnote{Absolute fidelity between the retrieved state and the ideal state}}&
			\textrm{Fidelity B\footnote{Relative fidelity between the retrieved state and the input state}}&
			\textrm{Classic limit}\\
			\colrule
			$\left|\Psi_A\right\rangle$ & $86.85\pm 4.00$\% & $89.13\pm 3.10$\% & $87.00$\%\\
			$\left|\Psi_B\right\rangle$ & $87.39\pm 1.73$\% & $88.83\pm 1.30$\% & $87.00$\%\\
		\end{tabular}
	\end{ruledtabular}
\end{table}

\section{Conclusion}
In this article, we first demonstrate the feasibility of an atomic ensemble working as a storage medium for transverse spatial modes, then discuss the two main decoherence mechanism in our experiemnt: atomic diffusion and unsynchronized evoulution of the stored coherence caused by ambient residual magneitc field. We briefly analyze the effect of atomic diffusion on the storage of an OAM qubit and an OAM qutrit, we find that an arbitrary OAM qudit has a phase structure that is intrinsically robust to atomic motion. We overcome the detrimental effect of ambient magnetic field by combining a guiding field and magnetically insensitive states. An atomic-based optical memory for OAM qutrits that can work in quantum region at least for a storage time of $t_s=400\mu s$ is realized in this work. To the best of our knowledge, this is two orders of magnitude longer than the previous work. 

Increasing the dimension of the stored qudits in further may be restricted by the penalty of dramatically increased number of basis that is needed for reconstructing the density operator, which put forward a stringent requirement on the long-term stability of the experimental system. What's more, as discussed before, the stored higher-dimensional transverse modes are expected to be more sensitive to atomic motion, therefore a lower ensemble temperature is appreciated. We can utilize techniques such as dark trap, optical molasses and evaporative cooling to achieve this goal. Since the conventional methods to constrain the motion of atoms such as dipole trap may be not suitable for storing transverse multimode because of the potential limitation arising from the achievable transverse size of an optical lattice, another fruitful research direction is to store multiple photonic degrees of freedom simultaneously\cite{spatialmultiplex,vectorbeamlaurat,vectorbeamye}.

\section{ackonwledgements}
This work was supported by National Key R\&D Program of China (2017YFA0304800), the National Nature Science Foundation of China (Grant Nos. 61722510, 11604322, 11934013), and the Innovation Fund from CAS, Anhui Initiative in Quantum Information Technologies (AHY020200), and the Youth Innovation Pro motion Association of Chinese Academy of Science under Grant No.2018490.

\bibliography{manuscript}
\end{document}